\documentclass{emulateapj}
\usepackage{graphicx}

\newcommand\mdot{\dot{M}}
\newcommand\game{\gamma_e}
\newcommand\rS{R_S}

\newcommand\K{{\rm\,K}}

\newcommand{\dv}[2]{\frac{d#1}{d#2}}

\newcommand{\be}{\begin{equation}}
\newcommand{\ee}{\end{equation}}
\newcommand{\bea}{\begin{eqnarray}}
\newcommand{\eea}{\end{eqnarray}}
\newcommand{\Par}{\parallel}
\shortauthors{Johnson \& Quataert}
\shorttitle{}
\begin{document}

\title{The Effects of Thermal Conduction on Radiatively-Inefficient Accretion Flows}

\author{Bryan M. Johnson and Eliot Quataert}

\affil{Astronomy Department,
601 Campbell Hall,
University of California at Berkeley,
Berkeley, CA 94720}

\begin{abstract}

We quantify the effects of electron thermal conduction on the
properties of hot accretion flows, under the assumption of spherical
symmetry. Electron heat conduction is important for low accretion rate
systems where the electron cooling time is longer than the conduction time
of the plasma, such as Sgr A* in the Galactic Center.  For accretion
flows with density profiles similar to the Bondi solution ($n[r]
\propto r^{-3/2}$), we show that heat conduction leads to super-virial
temperatures, implying that conduction significantly modifies the
dynamics of the accretion flow.  We then self-consistently solve for
the dynamics of spherical accretion in the presence of saturated
conduction and electron heating.  We find that the accretion rate onto
the central object can be reduced by $\sim 1-3$ orders of magnitude
relative to the canonical Bondi rate.  Electron conduction may thus be
an important ingredient in explaining the low radiative efficiencies
and low accretion rates inferred from observations of low-luminosity
galactic nuclei.  The solutions presented in this paper may also
describe the nonlinear saturation of the magnetothermal instability in
hot accretion flows.

\end{abstract}


\section{Introduction}\label{INTRO}

The conversion of gravitational binding energy of accreting matter
into radiation is thought to account for the high luminosities
observed from many compact objects (e.g., AGN; \citealt{lb69}). Most
of the time, however, systems that appear to harbor compact objects
(low luminosity AGN, X-ray binary systems in quiescence) do not
radiate at the levels one would expect from simple estimates of the
efficiency at which binding energy is converted to radiation. This
implies that these systems have low radiative efficiencies and/or low
accretion rates, and that the accreting gas may thus retain most of
its binding energy in the form of heat. The ultimate energy sink for
such hot radiatively inefficient accretion flows (RIAFs) remains an
area of active research. The energy may be primarily advected in with
the flow \citep{ny94}, or it may be directed into driving bulk
convective motion \citep{qg00a, nar00} or a global outflow
\citep{bb99}.  Numerical simulations favor the latter possibilities
(e.g., \citealt{ia99}; \citealt{sp01}; \citealt{hb02}).

Another possible outlet for the accretion energy is a heat flux due to
thermal conduction.  In order to maintain a hot accretion flow, the
timescale for electrons and protons to exchange energy by Coulomb
collisions exceeds the timescale for the plasma to flow into the
central black hole (e.g., Rees et al. 1982).  Thus the inflowing
plasma is collisionless and energy transport by heat conduction may be
dynamically important.  Direct observational estimates of the electron
mean-free path on scales resolved by {\it Chandra} in nearby galactic
nuclei -- including our Galactic Center -- give values $\sim 0.02-1.3$
times the gravitational radius of influence of the central black hole
($\sim 10^5-10^6$ Schwarzschild radii).  Thus the plasma is relatively
collisionless even at large distances from the black hole \citep{q04,
tm06}.  Under these conditions, the magnitude of the heat flux is
uncertain and depends on the geometry of the magnetic field and the
effective collision rate due to plasma waves and instabilities.  It is
plausible, however, that thermal conduction will occur at close to its
maximal saturated rate, with a heat flux $F \propto \rho v_e^3$, where
$\rho$ is the density and $v_e$ the electron thermal speed
\citep{cm77}. This appears to be true in the solar wind
\citep{salem03}, a well-studied example of a macroscopically
collisionless plasma.  For non-relativistic electrons, the ratio of
the inflow time to the electron conduction time in a hot accretion
flow may be as large as $\sim v_e/v_r$ (see equation [\ref{ENTe}]
below), where $v_r$ is the radial flow velocity. For subsonic inflow (with respect to the
ions), this ratio is $\gtrsim \sqrt{m_p/m_e} \sim 40$, highlighting
the possible importance of conduction in hot accretion flows.  In
addition, \citet{bal00} has shown that low collisionality conducting
plasmas are unstable to convective-like motions in the presence of an
outwardly decreasing temperature gradient (the MTI or magnetothermal
instability).  Although the interplay between the MTI and the
magnetorotational instability (MRI) is uncertain, it is plausible that
the MTI will lead to a dynamically important heat flux in hot
accretion flows.

Any significant heat flux from the inner (hotter) to the outer
(colder) parts of an accretion flow can act to reduce the inflow rate.
This is true if energy is transported by radiation (e.g., \citealt{owym76}), 
viscous stresses (e.g., \citealt{bb99}), convection (e.g., \citealt{qg00a}), 
or thermal conduction, the focus of this paper.  Gas in the outer region 
of a spherical accretion flow near the sonic point has a thermal energy 
comparable to its gravitational binding energy.  Additional heating will 
drive the temperature of the gas above the virial temperature. If the gas 
is unable to cool efficiently, this increase in temperature will lead to
a decrease in accretion rate by either decreasing the radius of the
sonic point or by driving a thermal outflow.

Motivated by these considerations, this paper assesses the importance
of thermal conduction using a simple one-dimensional (spherical) model
of hot accretion flows with saturated conduction.  \cite{tm06} have
carried out a related analysis and find that the accretion flow can
spontaneously produce thermal outflows driven in part by conduction
(as in early models of the solar wind).  Their analysis is
two-dimensional but self-similar in radius, whereas ours is
one-dimensional but does not assume self-similarity; we also treat
saturated heat conduction somewhat differently than they do (see
\S\ref{EE}).  Tanaka \& Menou's assumption of self-similarity enforces
a density profile that varies as $r^{-3/2}$, whereas simulations of
RIAFs consistently find density profiles shallower than this (e.g.,
\citealt{spb99}, \citealt{ia99}, \citealt{sp01}, \citealt{hb02},
\citealt{ina03}).

We begin in \S\ref{EE} by outlining the energy equation in the
presence of saturated heat conduction.  For completeness we consider
both the ion and electron energy equations, though our focus will be
on the electrons throughout this paper; we also quantify when electron
heat conduction is likely to be important as an energy transport
mechanism relative to radiative cooling.  We then present order of
magnitude arguments that demonstrate the importance of heat conduction
in hot accretion flows.  In particular, we solve for the electron
temperature in the presence of heat conduction under the simplifying
assumption that the density profile of the accretion flow is a fixed
power-law in radius (\S\ref{FDP}).  In \S \ref{section:bondi} we relax
the power-law density profile assumption of \S \ref{FDP} and
self-consistently solve for the density, temperature, and radial
velocity of the flow neglecting angular momentum; this is a version of
the classic \citet{bon52} accretion problem.  We summarize and discuss
the implications of our findings in \S \ref{DISC}.

\section{Energy Equations}\label{EE}

As noted in the Introduction, in a hot accretion flow the timescale
for electrons and ions to exchange energy by Coulomb collisions
exceeds the timescale for the plasma to flow into the central black
hole.  Thus the flow likely develops a two-temperature structure,
although the extent to which the particle species are coupled through
other processes such as wave-particle interactions is unclear. In
general, then, the energy of the electrons and ions must be tracked
separately.  In spherical symmetry, the electron and ion energy
equations can be written as

\be\label{ENTe} -v_r n T_e \dv{s_e}{r} = q_e^+ - \frac{1}{r^2}
\dv{}{r}\left(r^2 Q_e \right) - q_e^-, \ee and \be\label{ENTi} -v_r n
T_i \dv{s_i}{r} = q_i^+ - \frac{1}{r^2} \dv{}{r}\left(r^2 Q_i \right)
\ee where $v_r \equiv -dr/dt$ is the radial velocity (defined here to
be positive inward), $n$ is the particle number density ($n_e = n_i
\equiv n$ by charge neutrality, where the subscripts $e$ and $i$
denote properties of the electron and ion fluids, respectively), $s$
is the specific entropy, $q^+$ is the heating rate per unit volume,
and $q^-_e$ is the electron cooling rate per unit volume.\footnote{We
work in units in which Boltzmann's constant $k_B \equiv 1$ so that
temperature has units of energy, although we express our numerical
results in Kelvin.}  In equations (\ref{ENTe}) and (\ref{ENTi}) we
have neglected energy exchange between ions and electrons (e.g., via
Coulomb collisions) and ion radiative cooling.  The last term on the
right-hand side of equations (\ref{ENTe}) and (\ref{ENTi}) is the
divergence of the radial conductive heat flux \be\label{QR} Q_{e,i}
\equiv -\kappa_{e,i} \dv{T_{e,i}}{r}. \label{flux} \ee

Because the accreting plasma is collisionless, we model the thermal
conductivity $\kappa$ using a saturated thermal conductivity
\citep{cm77}: \be \label{KAPPA} \kappa_e \equiv \alpha_c v_c n r, \ee where $\alpha_c$
is a dimensionless parameter that measures the strength of the
conduction and \be v_c \equiv \frac{c v_e}{c + v_e} \simeq \left\{
\begin{array}{cc}
v_e & {\rm for} \; v_e \ll c \\ c & {\rm for} \; v_e \gg c \\
\end{array}
\right.  \label{vc} \ee is a function that interpolates smoothly
between the electron thermal speed at low
temperatures and the speed of light $c$ as the electrons become
relativistic. For the ions, which are always non-relativistic,
$\kappa_i \equiv \alpha_c v_i n r$ where $v_i$ is the ion thermal
velocity.  We use a slightly different form for the saturated flux
than \cite{cm77}. Their flux is proportional to density and
temperature, while we take the flux to be proportional to the
temperature gradient.  For our problem, the temperature and density
are boundary conditions at large radii; the \cite{cm77} prescription
would lead to the heat flux being specified by these boundary
conditions, which is unphysical since we want to be able to solve for
the heat flux rather than have it predetermined.  We note, however,
that our assumption of $Q \propto dT_e/dr$, while physically
reasonable, cannot be rigorously derived in the collisionless limit,
where the heat flux depends on a non-local integral of the electron
temperature along magnetic field lines (see, e.g., \S VII of Snyder et
al. 1997).\footnote{If wave-particle scattering limits the mean free
path of particles to be $\lesssim r$ at any radius, then $Q \propto
dT_e/dr$ is again well-justified.}

Assuming that a fraction $f_e$ ($f_i$) of the available accretion
energy goes into heating the electrons (ions), the electron and ion
heating rates per unit volume are given by \be q^+_{e,i} = f_{e,i}
\frac{G M \mdot}{4 \pi r^4}. \label{eheat} \ee where $f_i = 1 - f_e$.
Integrating the total (electron + ion) volumetric heating rate over
all radii yields a net heating rate of $GM \dot M /R_{in}$, where
$R_{in}$ is the inner radius of the flow.  We set $R_{in} = 3 \rS = 6
GM/c^2$ throughout this paper so that the total heating rate is
$\approx 0.16 \dot M c^2$.  Note that even for spherical accretion
(i.e., in the absence of differential rotation and the MRI), a
significant fraction of the gravitational potential energy of the
inflowing gas may be converted into heat.  In particular, even
initially weak magnetic fields are compressed by flux freezing until
they become strong, leading to reconnection and significant heating
\citep{in02}.

Because $v_e \gg v_i$ unless $T_e \lesssim 10^{-3} \, T_i$, electron
heat conduction tends to be more important than ion heat conduction.
Our focus in this paper will thus be on electron heat conduction.  We
note, however, that electron heat conduction is important only when
electron radiative losses are relatively small.  Otherwise, most of
the energy supplied to the electrons is radiated away and there is
little energy available to redistribute by conduction.  Throughout
this paper we thus set $q_e^- = 0$ in equation (\ref{ENTe}).  This is
valid only when the cooling time of the electrons in the accretion
flow is long compared to the conduction time, $r/v_c$.  At the low
accretion rates where this assumption is appropriate (see below), and
inside $\sim 10^3 \rS$ where the electrons are relativistic,
synchrotron cooling is the dominant cooling mechanism (e.g.,
\citealt{qn99}).  Synchrotron cooling for thermal electrons at
temperature $T_e$ can be neglected only for \be\label{MDUL} {\dot M
\over \dot M_{\rm EDD}} \lesssim 10^{-6} \left(\frac{\beta}{10}\right)
\left(\frac{\alpha_c}{0.1}\right) \left(\frac{\alpha}{0.1}\right)
\left(\frac{T_e}{10^{11} \K}\right)^{-1} \left(\frac{r}{3
\rS}\right)^{3/2}, \ee where $\mdot_{\rm EDD} \equiv L_{\rm EDD}/(0.1
c^2)$ is the Eddington accretion rate, the radial velocity is assumed
to be a fraction $\alpha$ of the free-fall velocity, the magnetic
field has been normalized to the virial temperature via $\beta \equiv
n T_i/(B^2/8\pi)$, the average Lorentz factor for a thermal
distribution of electrons is given by $3 T_e/(m_e c^2)$, and $v_c
\approx c$ for relativistic electrons. Expression (\ref{MDUL}) is
considerably more restrictive than the requirement that $\dot M
\lesssim \alpha^2 \dot M_{\rm EDD}$ for Coulomb collisions to be
unimportant, which is necessary to maintain a two-temperature flow
\citep{rpbb82}. There is therefore a large range of accretion rates
where a RIAF is possible but where electron heat conduction is
unimportant because the electrons radiate away most of their
energy.\footnote{Note that at high accretion rates electron conduction
could still be important at radii $\gtrsim 10^3 \rS$ where the
electrons are non-relativistic and synchrotron losses are small.} Our
focus here is on very low accretion-rate systems where electron
radiative losses are small.  This includes Sgr A* in the Galactic
Center and other very low-luminosity AGN and X-ray binaries in
quiescence. Ion heat conduction may be important at higher accretion
rates, but this is harder to assess because the ion conduction time is
comparable to the inflow time.

We note that in their treatment of heat conduction in RIAFs,
\cite{tm06} use a single energy equation, as we do. Interpreted as
electron conduction, their results-- like ours-- are only relevant for
very low accretion rates. Interpreted as ion conduction, their
$\phi_S$ should be $\lesssim 0.01$ (and their $\lambda_0 \lesssim
0.1$) since using the ion thermal speed in the conductivity introduces
an additional factor of $\sqrt{m_e/m_p}$. For these values, their
solutions show that ion conduction has only a modest effect on the
properties of RIAFs.

The appropriate values for $\alpha_c$ and $f_e$ are set by complicated
physical processes that are not fully understood.  The electron
heating parameter must satisfy $f_e \lesssim 1$ on energetic grounds.
For reasons explained in, e.g., \citet{qg99} and \citet{qua03}, we
suspect that the electrons likely receive a significant fraction of
the available gravitational potential energy in RIAFs, so that $f_e
\sim 0.1-1$ is reasonable.  The dimensionless conductivity $\alpha_c$
must also satisfy $\alpha_c \lesssim 1$.  We show solutions with
$\alpha_c \gtrsim 1$, however, in order to illustrate the change in
dynamics as the inflowing plasma becomes isothermal. In the Appendix
we place an upper limit on the ion conductivity of $\alpha_c \lesssim
0.1$ based on pitch angle scattering by instabilities generated by the
nonlinear evolution of the MRI in a collisionless plasma.  This argues
against the importance of ion conduction.  The corresponding
constraint on the electron conductivity is much weaker
(eq. \ref{elimit}). 

With equations (\ref{ENTe})-(\ref{eheat}) describing the electron
energetics, we now proceed to assess the importance of heat conduction
in RIAFs by considering a couple of model problems.

\section{Order-of-Magnitude Considerations}\label{FDP}

To quantitatively motivate the importance of thermal conduction, we
consider a simple model in which a fraction $f_e$ of the accretion
energy $(0.1 \mdot c^2$) is carried out to large radii by a saturated
conductive heat flux with $\alpha_c \sim 1$, in which case $0.1 f_e
\mdot c^2 \approx 4\pi \rho r^2 v_e^3 (m_e/m_p)$.
For constant $\mdot$, this implies \be \frac{T_e}{T_{vir}} \sim
\left(0.1 f_e \sqrt{\frac{m_e}{m_p}} \, \frac{r}{\rS}\right)^{2/3}
\sim 30 \left(f_e \, \frac{r}{10^5 \rS}\right)^{2/3},
\label{trbondi}\ee where we have
assumed that the electrons are non-relativistic, which is true at
large radii, $r/\rS \gtrsim 10^3$.  Equation (\ref{trbondi}) shows that
if even a small fraction of the available energy is transported
outwards by conduction ($f_e \gtrsim 0.01$), the heat flux can unbind
the gas at large radii, thus significantly changing the flow dynamics.

The above analysis assumes that $\mdot$ is constant and thus that the
density profile of the flow is given by $\rho \propto r^{-3/2}$, the
canonical result for spherical accretion.  Models of rotating
radiatively inefficient accretion flows show, however, that the
dynamics of the flow can be quite different from that of spherical
accretion.  In particular, convection (\citealt{spb99,qg00a, nar00})
and/or thermally driven winds (\citealt{bb99,sp01,hb02}) can strongly
suppress the rate of mass accretion onto the central black hole.  A
simple model for such a flow is one in which the density profile is
given by $n \propto r^{-p}$, so that $\mdot \propto r^{3/2-p}$ is
steady in time but varies with radius.  Numerical simulations favor $p
\approx 1/2-1$.  For $p < 3/2$, the mass accretion rate decreases at
small radii, and there is {\it less} energy available to be conducted
out to large radii.  In this case, energy transport by conduction has
less of an effect on the dynamics of the flow.  To quantify this, we
again assume that a fraction $f_e$ of the accretion energy at small
radii $(0.1 \mdot[\rS] c^2$) is carried out to large radii by a
saturated conductive heat flux; allowing for $\mdot \propto
r^{3/2-p}$, the temperature at large radii is given by \be
\frac{T_e}{T_{vir}} \sim \left(\frac{0.1 f_e}{p - 1/2}
\sqrt{\frac{m_e}{m_p}} \left[\frac{r}{\rS}\right]^{p-1/2}\right)^{2/3}
\label{trp}\ee for $p > 1/2$ and \be \label{P12} \frac{T_e}{T_{vir}}
\sim \left(0.1 f_e \sqrt{\frac{m_e}{m_p}} \ln
\left[\frac{r}{\rS}\right]\right) ^{2/3}
\label{trp0.5}\ee for $p = 1/2$.  For $p = 1/2$, equation
(\ref{trp0.5}) shows that the electron temperature required to carry
the heat flux out is always sub-virial and thus conduction is not
dynamically important.  More generally, from equation (\ref{trp}), and
assuming a fiducial outer-radius of $r \sim 10^5 \rS$, we find that
the electron temperature is super-virial for $p \gtrsim 1$.  

To expand on the above analysis, we solve the full electron energy
equation (eq. [\ref{ENTe}]) using a standard relaxation algorithm
\citep{nr92} for a fixed density profile $n \propto r^{-p}$. We take
the the radial velocity to be given by \be v_r \equiv \alpha
\left(\frac{H}{r}\right)^2 v_\phi \simeq \alpha \sqrt{\frac{GM}{r}},
\label{vr} \ee where $H \simeq r$ is the vertical scale-height for a hot 
accretion flow, $v_\phi = r\Omega$ is the rotational velocity and
$\Omega = \sqrt{GM/r^3}$ is the Keplerian rotational frequency.  There
are two boundary conditions for this problem since we are specifying
both the radial velocity and density profiles. A reasonable set of
boundary conditions is to specify the temperature at the outer radius
and to require that $dT_e/dr|_{R_{in}} = 0$, i.e., that the conductive
heat flux is zero at the inner radius (so that there is not a source
of energy at the horizon). Our results are all insensitive to the
precise inner boundary condition (for modest values of
$dT_e/dr|_{R_{in}}$).  The parameters that determine the electron
temperature in this model are $\alpha_c/\alpha$, $p$, and $f_e$.

\begin{figure}
\plotone{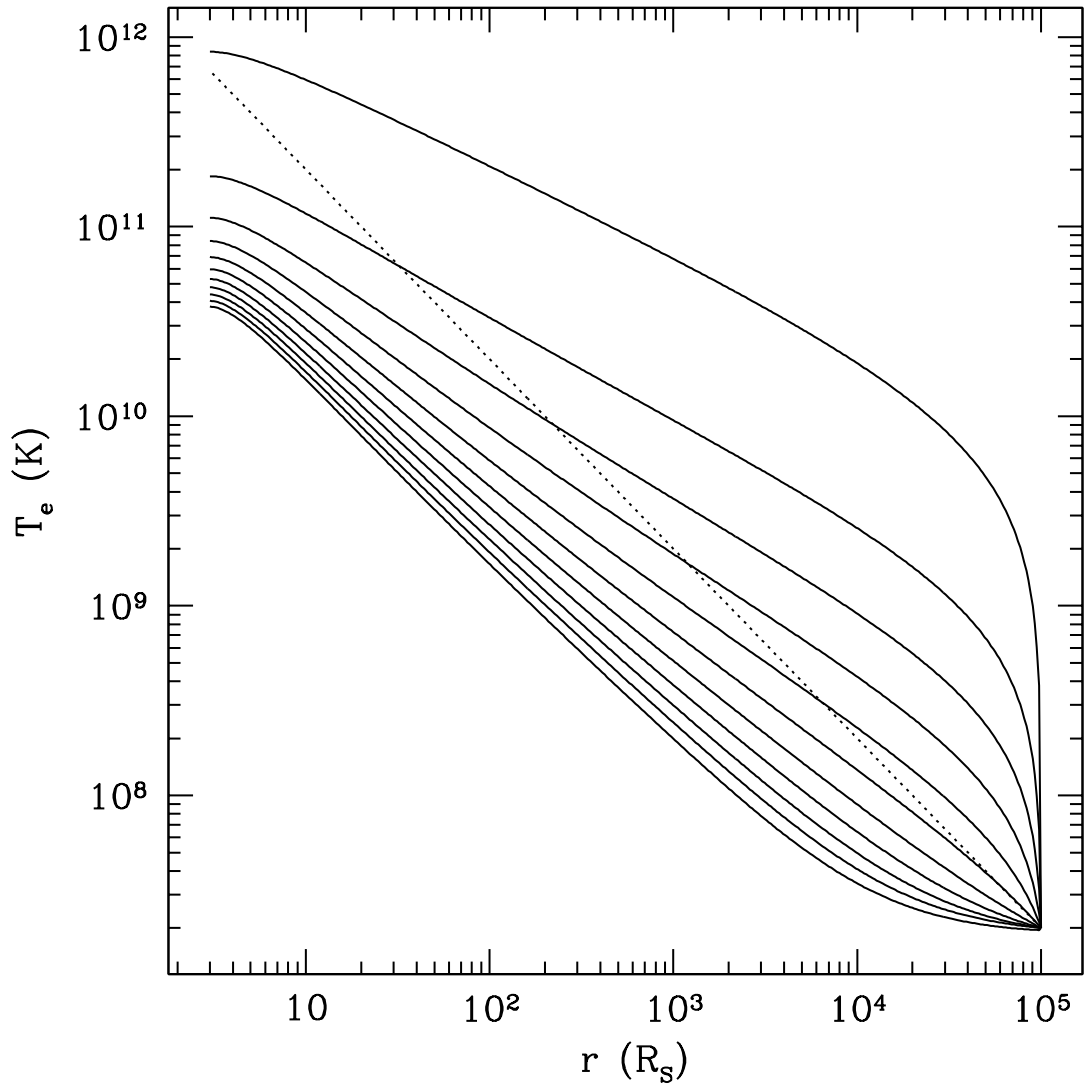}
\caption{Electron temperature for a fixed density profile with $n
\propto r^{-p}$, with $p$ varying from $0.5$ to $1.5$
in increments of $0.1$ (from bottom to top). The dotted line is the
virial temperature $T_{vir} \propto r^{-1}$. Values for the other
parameters are $\alpha_c/\alpha = 1$ and $f_e = 0.1$. For $p \gtrsim
1$, these solutions are unphysical because the electron temperature is
super-virial.}
\label{f1}
\end{figure}

Figure~\ref{f1} is a plot of $T_e(r)$ for different choices for the
density power-law index $p$. The values for the other parameters are
$\alpha_c/\alpha = 1$, $f_e = 0.1$ and an outer temperature $T_{out} =
2\times 10^7 \K$ (the temperature observed in the Galactic Center;
Baganoff et al. 2003).  Figure \ref{f1} shows that larger values of
$p$ lead to larger electron temperatures (see also
eqs. [\ref{trbondi}]-[\ref{trp0.5}]).  This is because larger values
of $p$ correspond to larger accretion rates and electron heating rates
at small radii.  In the presence of significant heat conduction some
of the energy supplied to the electrons at small radii is conducted to
large radii (for $\alpha_c \sim \alpha$, the heat conduction time is
$\sim v_c/v_\phi \gtrsim 1$ times smaller than the inflow time; thus
most of the energy supplied to the electrons is conducted to large
radii, as assumed in the order of magnitude estimates above).  Larger
values of $p$ thus correspond to a larger heat flux at large radii,
and a larger electron temperature is needed to carry this heat flux
(eqs. [\ref{flux}]-[\ref{vc}] imply $Q_r \propto T_e^{3/2}$ at large
radii where the electrons are non-relativistic).  For sufficiently
large values of $p \gtrsim 1$ the solutions shown in Figure~\ref{f1} are
unphysical: the electron temperature becomes super-virial, indicating
that the electron pressure gradient is dynamically important and that
the electrons are unbound from the system (see also eq. [\ref{trp}]).

Using solutions with $p = 1$ and $p = 1/2$ as reference calculations
(values motivated in part by simulations of RIAFs), we explore how the
temperature profile varies as a function of $\alpha_c/\alpha$ and
$f_e$.  These calculations are shown in Figures~\ref{f2} and
\ref{f3}. The trends in these figures can all be understood relatively
easily.  Larger values of $f_e$ correspond to larger electron heating
rates and thus lead to higher electron temperatures.  Larger values of
$\alpha_c/\alpha$, on the other hand, correspond to very efficient
heat conduction that drives the electron temperature profile towards
isothermality.  Note that in the case of $p = 1/2$, the electron
temperature is never super-virial (Fig.~\ref{f3}), consistent with our
order-of-magnitude estimate in equation (\ref{P12}).

\section{Spherical Accretion with Conduction}\label{section:bondi}

The fact that the electron temperature calculated in the previous
section is sometimes super-virial implies that conduction can
significantly influence the dynamics of the accretion flow;
this motivates the investigation of this subsection, in which we solve
for the density and velocity profiles self-consistently.  We consider
a simple model of a hot accretion flow, assuming spherical symmetry,
constant $\mdot$, and neglecting angular momentum.

With the above assumptions the basic dynamical equations for the
density and radial velocity are \be\label{MASS} \mdot = 4\pi r^2 \rho
v_r \ee \be\label{MOM} v_r \dv{v_r}{r} + \frac{1}{\rho} \dv{P_e}{r} +
\frac{G M}{\left(r - \rS\right)^2} = 0, \ee where $P_e$ is the electron pressure 
and we have used the \cite{pw80} gravitational potential to ensure that 
all the solutions have a sonic radius outside the horizon.  In
addition to equations (\ref{MASS}) and (\ref{MOM}) we solve equation
(\ref{ENTe}) for the electron temperature, which can also be written
as \be\label{ENT2} n v_r \dv{}{r}\left(\frac{T_e}{\game - 1}\right) -
v_r T_e \dv{n}{r} + \frac{f_e G M m_p n v_r}{r^2} +
\frac{\alpha_c}{r^2} \dv{}{r}\left(r^3 n v_c \dv{T_e}{r}\right) = 0,
\ee where the entropy gradient has been rewritten in terms of density
and temperature gradients and the internal energy has been expressed
in terms of an effective adiabatic index \be \game \equiv
\left(\frac{10}{3}\right)\left(\frac{1 + 2 v_e^2/c^2}{2 + 5 v_e^2/c^2}\right)
\simeq \left\{
\begin{array}{cc}
5/3 & {\rm for} \; v_e \ll c \\ 4/3 & {\rm for} \; v_e \gg c \\
\end{array}
\right.  \ee that interpolates smoothly between nonrelativistic and
relativistic energies.  Equations (\ref{MASS})-(\ref{ENT2}) are the
\cite{bon52} accretion problem except that there is now explicit
heating and an additional dependent variable ($dT_e/dr$) due to the
conductive flux term in equation (\ref{ENT2}).  We note that
\cite{gru98} has also considered Bondi accretion in the presence of
heat conduction, although his form of the conductive heat flux is quite
different from ours and leads to different results.

In general, the pressure in equation (\ref{MOM}) is given by the sum
of the electron and ion gas pressures (radiation pressure is
negligible since the plasma is optically thin and we do not include
the effects of magnetic fields) and one needs to solve equation
(\ref{ENTi}) for the ions in addition to equation (\ref{ENT2}) for the
electrons. Since our goal is to understand the effects of conduction
on the solution, however, it is reasonable to neglect the ion pressure
in equation (\ref{MOM}) and focus solely on the electron energetics.
The ion temperature in hot accretion flows is approximately virial,
$T_i \propto 1/r$, where the constant of proportionality depends upon
the assumption one makes about the fraction $f_i$ of accretion energy
that goes into heating the ions.  In \S \ref{DISC}, we briefly discuss
the effect of including the ion energy equation in our solutions.

Eliminating the density as a dependent variable in favor of the radial
velocity, the momentum equation can be expressed as \be\label{VR} \left(v_r^2 -
c_s^2\right)\frac{1}{v_r}\dv{v_r}{r} = \frac{2 c_s^2}{r} - \dv{c_s^2}{r}
- \frac{G M}{(r-\rS)^2}, \ee where $c_s^2 \equiv P_e/\rho$ is the
isothermal sound speed.  Equation (\ref{VR}) is singular when the
factor in parentheses on the left-hand side is zero.  By analogy with
the Bondi problem, we expect a family of subsonic solutions as well as
a unique solution that makes a transition between subsonic and
supersonic flow at an intermediate radius $R_{sonic}$ (the sonic
point). The solution that passes through the sonic point is selected
by setting both the left-hand and right-hand sides of equation
(\ref{VR}) to zero at $R_{sonic}$.  Specifying two additional
boundary conditions over-constrains the problem, so that $R_{sonic}$
(and therefore $\mdot$) is uniquely determined as an eigenvalue. As in
the previous section, the additional two boundary conditions are the
temperature at the outer radius $T_{out}$ and the zero-flux condition
($dT_e/dr = 0$) at the inner radius.  Note that the problem solved in
this section is thus a 3 point boundary value problem, with boundary
conditions at the inner radius, sonic point, and outer radius.  We
solve this problem with a relaxation algorithm.

\begin{figure}
\plotone{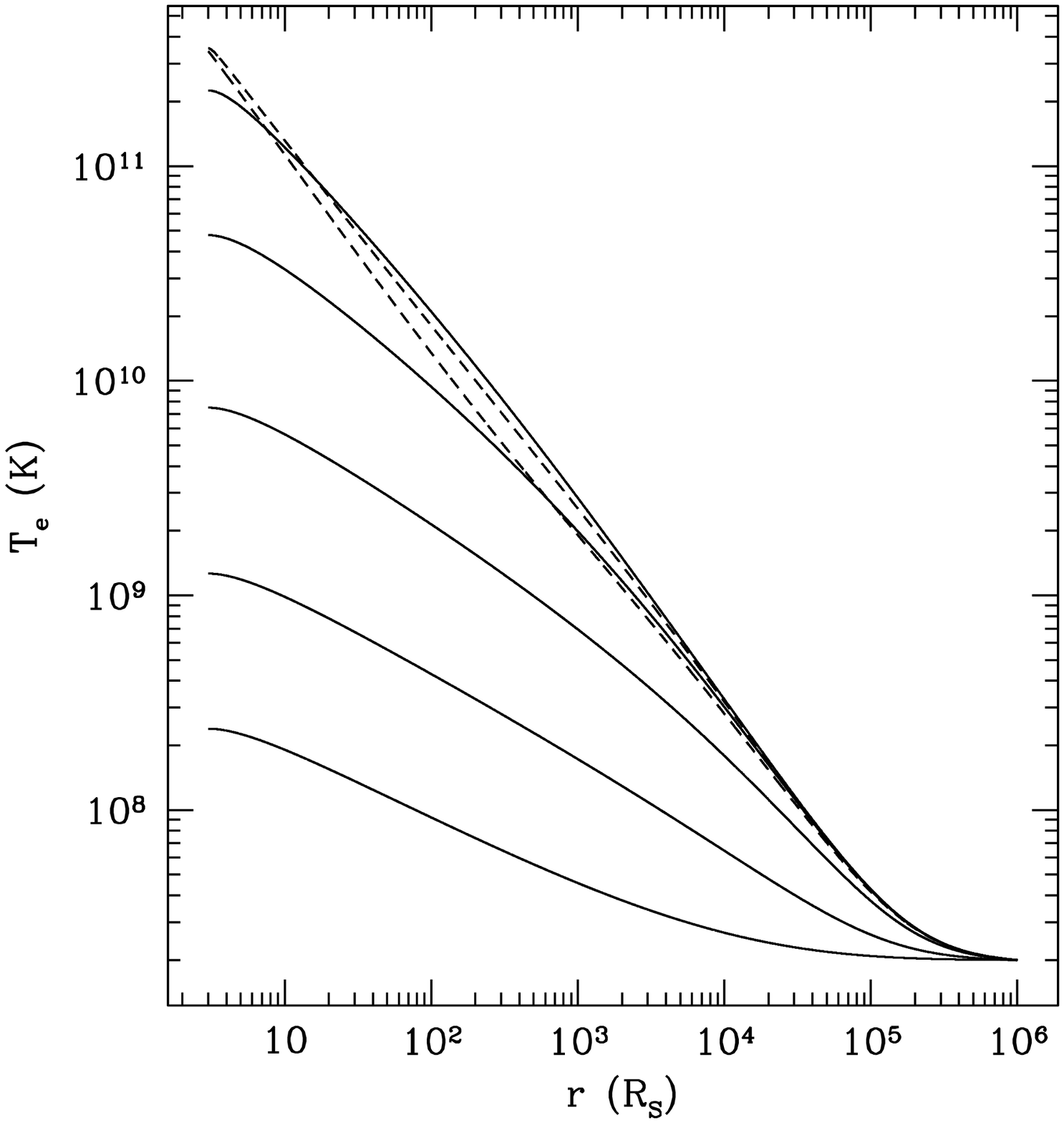}
\caption{Radial temperature profile for the solution of the Bondi
accretion problem with heating and conduction, for $f_e = 0.3$.  The
solid curves correspond to $\alpha_c$ varying from $1$ to $10^4$ by
factors of 10 (from top to bottom).  The dashed curves are for
$\alpha_c = 10^{-2}$ (top) and $0.1$ (bottom).}
\label{f4}
\end{figure}

\begin{figure}
\plotone{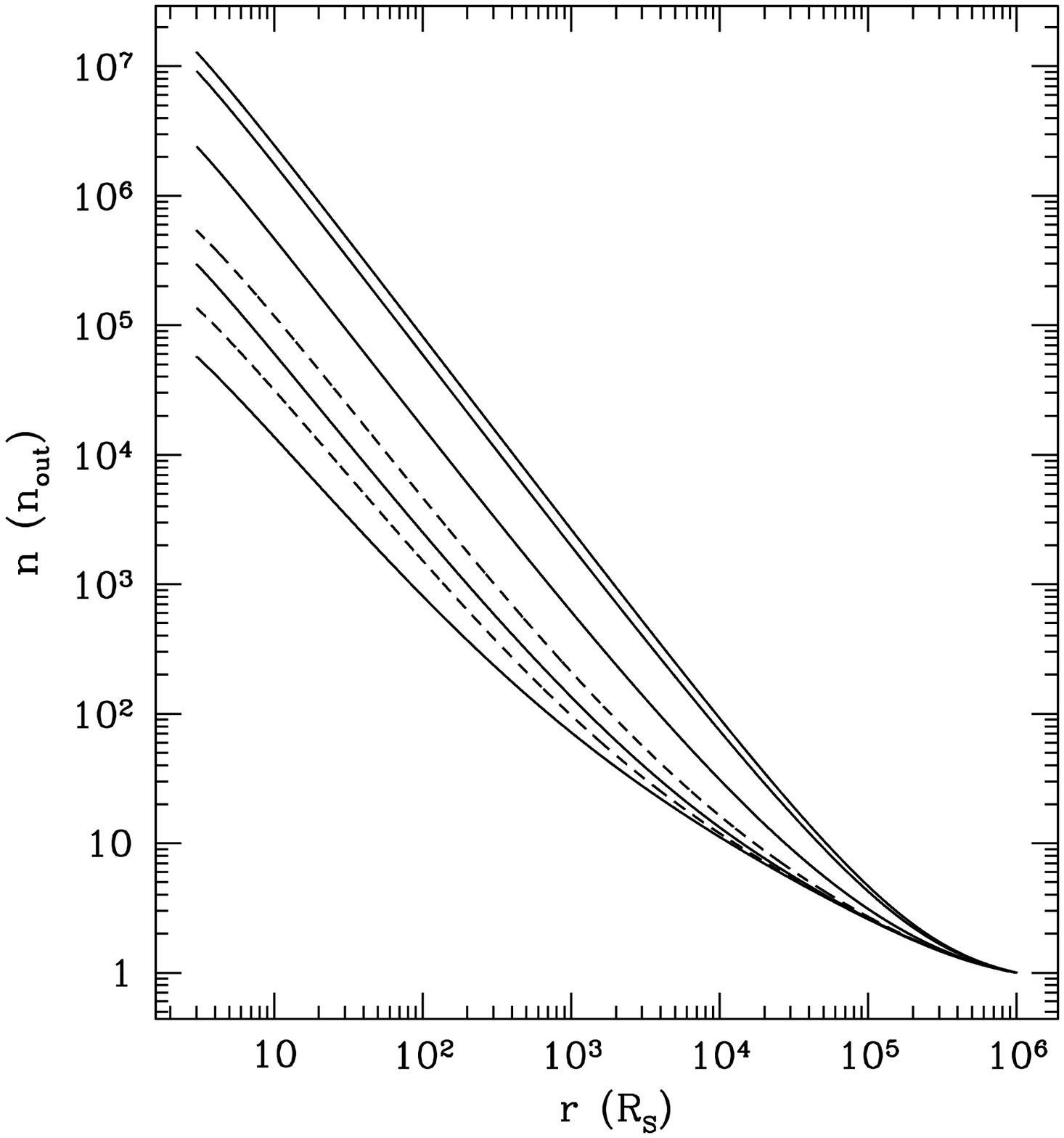}
\caption{Radial density profile for the solution of the Bondi
accretion problem with heating and conduction, for $f_e = 0.3$.  The
solid curves correspond to $\alpha_c$ varying from $1$ to $10^4$ by
factors of 10 (from bottom to top).  The dashed curves are for
$\alpha_c = 10^{-2}$ (top) and $0.1$ (bottom).}
\label{f5}
\end{figure}

\begin{figure}
\plotone{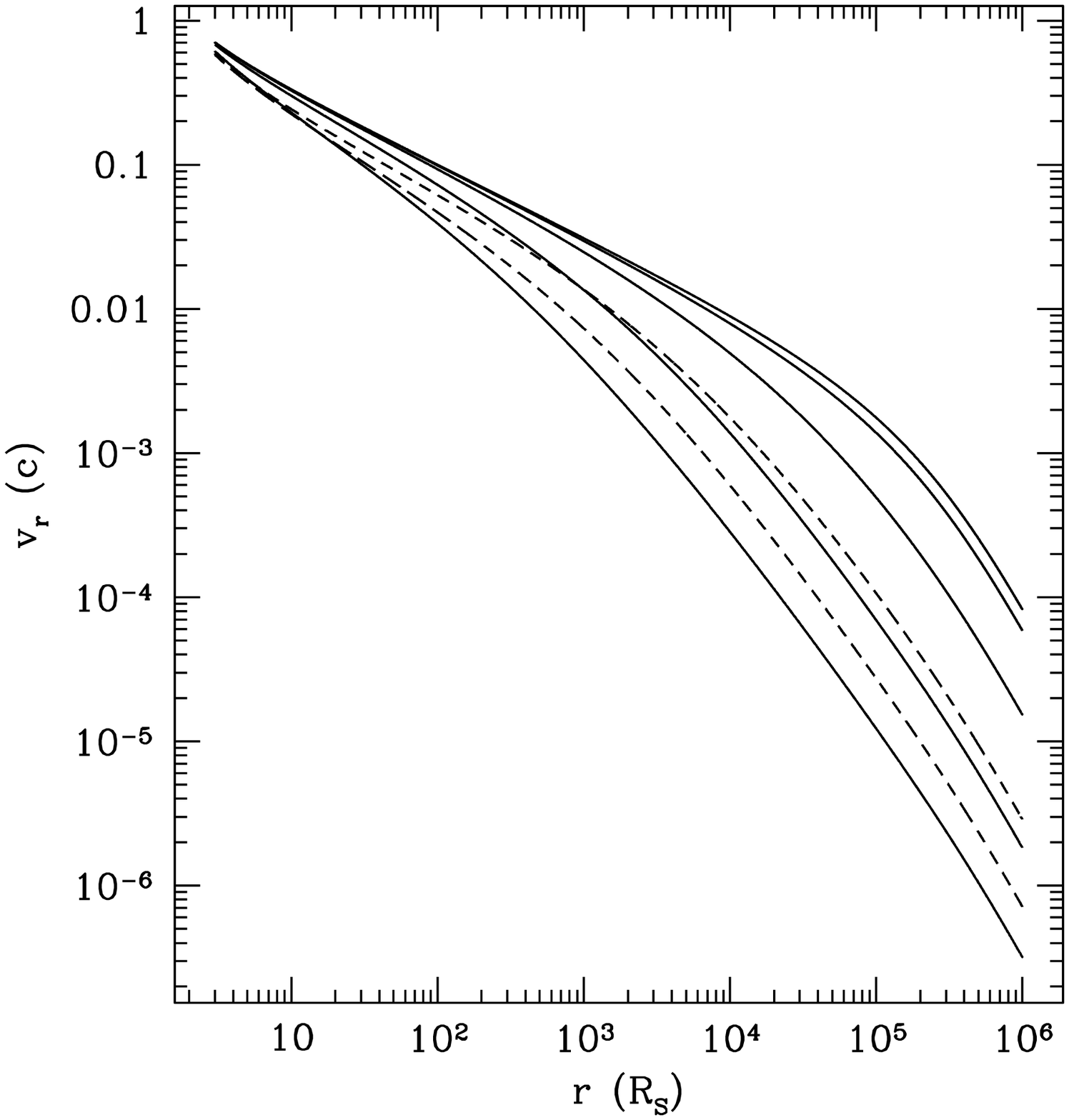}
\caption{Radial velocity profile for the solution of the Bondi
accretion problem with heating and conduction, for $f_e = 0.3$.  The
solid curves correspond to $\alpha_c$ varying from $1$ to $10^4$ by
factors of 10 (from bottom to top).  The dashed curves are for
$\alpha_c = 10^{-2}$ (top) and $0.1$ (bottom).}
\label{f6}
\end{figure}

\begin{figure}
\plotone{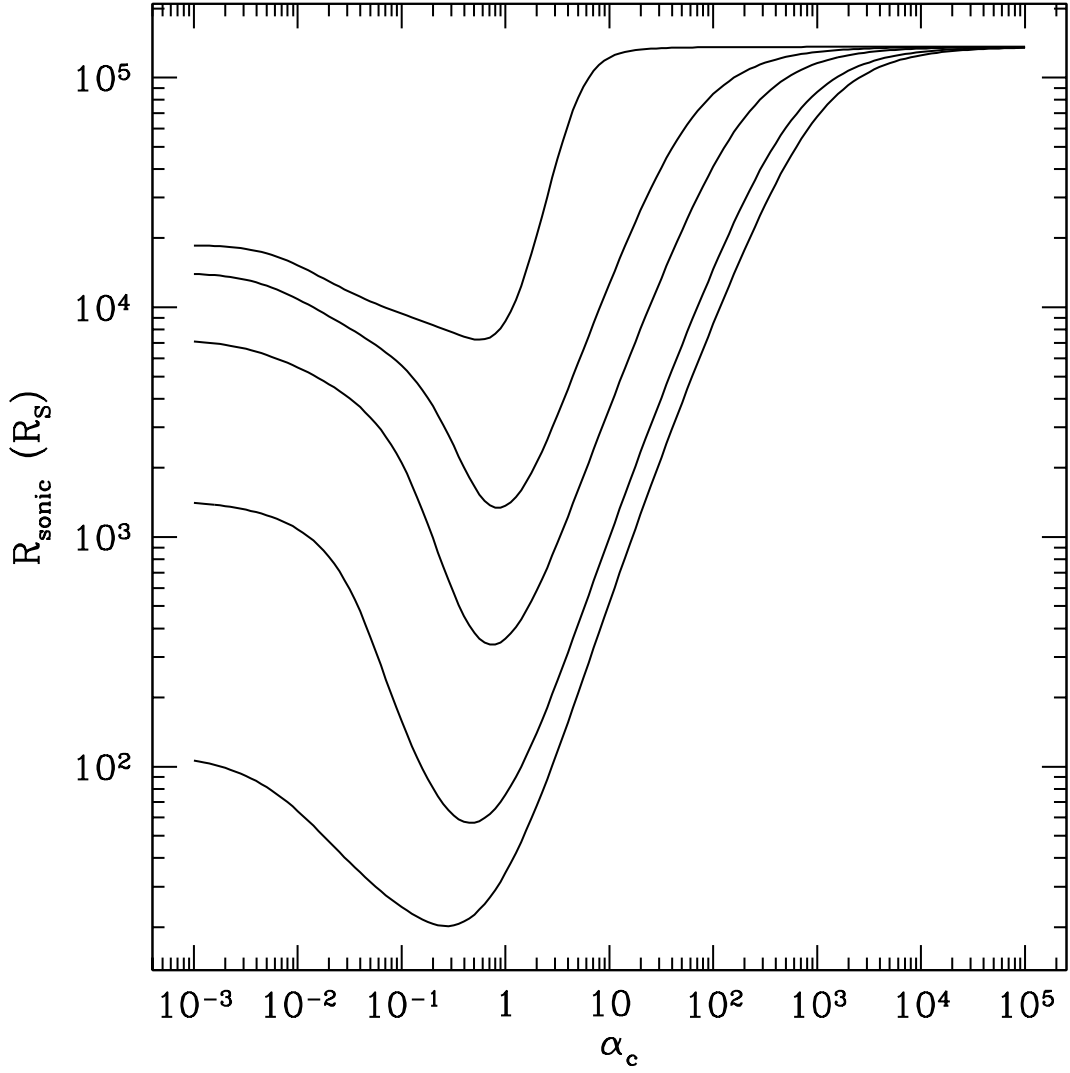}
\caption{ Sonic radius (in units of the Schwarzschild radius $\rS$) as
a function of $\alpha_c$ for $f_e = 0$ (top), $0.03$,
$0.1$, $0.3$ and $0.5$ (bottom).  }
\label{f7}
\end{figure}

\begin{figure}
\plotone{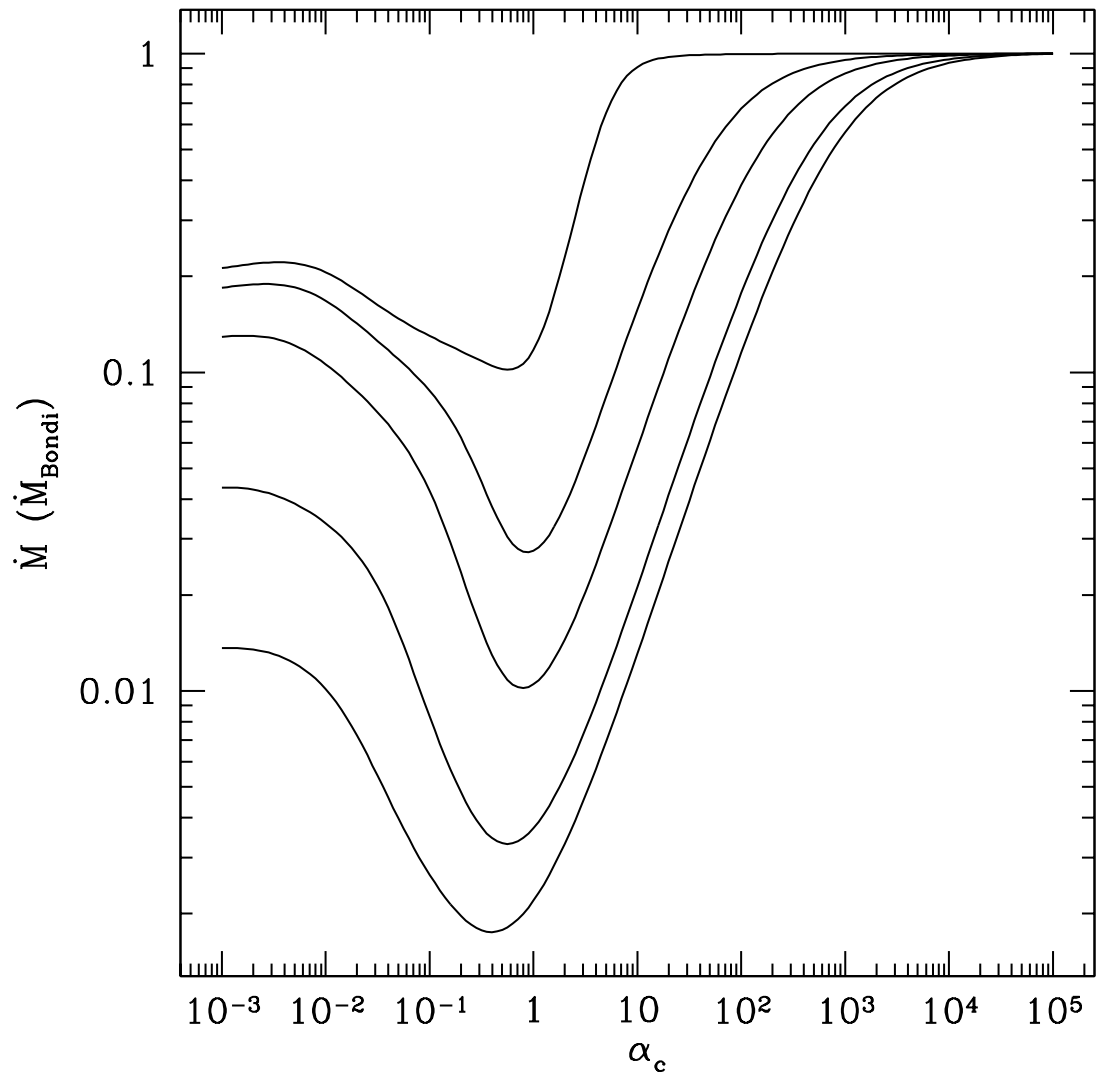}
\caption{
Accretion rate (in units of the isothermal Bondi rate) as a function of $\alpha_c$ for $f_e = 0$ (top), $0.03$, $0.1$, $0.3$ and $0.5$ (bottom). 
}
\label{f8}
\end{figure}

Figures~\ref{f4}, \ref{f5} and \ref{f6} show the temperature, density
and velocity profiles for a set of calculations with $f_e = 0.3$,
$T_{out} = 2 \times 10^7 \K$ and various values of $\alpha_c$.  Our
choice of $T_{out}$ corresponds to a gravitational sphere of influence
of the black hole of $GM/c_s^2|_{R_{out}} \approx 10^5 \rS$. As in
Figures~\ref{f2} and \ref{f3}, large values of $\alpha_c$ drive the
electron temperature to be isothermal. The electron temperature is
maximized (and the density at small radii minimized) for $\alpha_c
\sim 1$. The differences relative to the standard Bondi problem are
quantified more clearly in Figures~\ref{f7} and \ref{f8}, which show
the sonic radius and accretion rate as a function of $\alpha_c$ for
several values of $f_e$.  In all of our calculations, the standard
Bondi problem is recovered as $\alpha_c \rightarrow 0$ and $f_e
\rightarrow 0$, though note that we are employing an adiabatic index
that varies with radius so the $\alpha_c \rightarrow 0$ solution does
not correspond to a constant $\gamma$ as in the Bondi problem. In the
limit of infinite conductivity, $\alpha_c \gg 1$, the electron
temperature profile becomes isothermal, independent of $f_e$, and the
problem reduces to the standard Bondi problem with $\gamma = 1$.

For intermediate values of $\alpha_c \sim 1$, however, the solution
differs significantly from the canonical Bondi solution.  In
particular, Figure \ref{f7} shows that for moderate values of $f_e$
and $\alpha_c \sim 0.1-1$, the sonic point moves in from $\sim 10^5 \,
\rS$ to $\sim 10-100 \, \rS$.  The accretion rate is correspondingly
reduced relative to the adiabatic or isothermal Bondi rate by $\sim
1-3$ orders of magnitude (Fig. \ref{f8}).  Physically, this occurs
because accretion at the Bondi rate would imply so much heat
conduction to large radii that the electron temperature would be
super-virial, as found in the previous section.  To maintain $T_e
\lesssim T_{vir}$ requires that the accretion rate decrease so that
less energy is conducted out to large radii.  This occurs via a
decrease in the radial velocity (Fig. \ref{f6}) relative to free fall
and a flattening of the radial density profile relative to the
$r^{-3/2}$ Bondi solution (Fig. \ref{f5}).

The fact that the accretion rate for nonzero $f_e$ differs from the
adiabatic Bondi rate at low $\alpha_c$ can be understood as
follows.  In the outer, subsonic region, the momentum equation is
dominated by hydrostatic balance: \be\label{SUB}
\dv{\left(nT_e\right)}{r} + \frac{G M m_p n}{r^2} = 0.  \ee For
$\alpha_c = 0$, the energy equation is given by \be\label{AC0} n
\dv{}{r}\left(\frac{T_e}{\game - 1}\right) - T_e \dv{n}{r} + \frac{
f_e G M m_p n}{r^2} = 0.  \ee Combining these two relations gives \be
\dv{}{r}\left[\frac{\game}{\game - 1}T_e - \frac{GMm_p}{r} - f_e
\frac{GMm_p}{r}\right] = 0, \ee which for the form of $\game$ we
have chosen gives a quadratic relation for $T_e$. For constant $\gamma$, 
$T_e \propto 1/r$ and $n \propto r^{-p}$,
with \be\label{PEFF} p = \frac{1 - f_e\left(\gamma -
1\right)}{\left(1+ f_e\right)\left(\gamma - 1\right)}.  \ee The
assumption of hydrostatic equilibrium breaks down for
$p > 3/2$ (the free-fall value), or \be\label{GAMC} \gamma < 
\gamma_{crit} \equiv \frac{5\left(1 + f_e\right)}{3\left(1 + 5f_e/3\right)}. \ee 
For $\gamma < \gamma_{crit}$ a supersonic free-fall solution is obtainable, 
while for larger $\gamma$ it is not. Since $\gamma_{crit} < 5/3$ for nonzero 
$f_e$, the assumption of hydrostatic equilibrium holds for a larger range of 
radii than it would in the absence of heating
(until the temperature increases and $\game$ decreases enough to 
violate condition [\ref{GAMC}]). There is a corresponding decrease 
in the sonic radius and accretion rate even for $\alpha_c \rightarrow 0$, 
as can be seen in Figures~\ref{f7} and \ref{f8}.
Also, since $\game > 4/3$, equation (\ref{GAMC}) indicates that
only solutions with $f_e < 0.6$ will be able to make a transition
to supersonic flow as $\alpha_c \rightarrow 0$ (in a Newtonian 
potential). For larger values of $f_e$ the flow is in approximate
hydrostatic equilibrium down to very small radii, where the
singularity in the pseudo-Newtonian potential forces a sonic
transition. 

\section{Discussion}\label{DISC}

\begin{figure}
\plotone{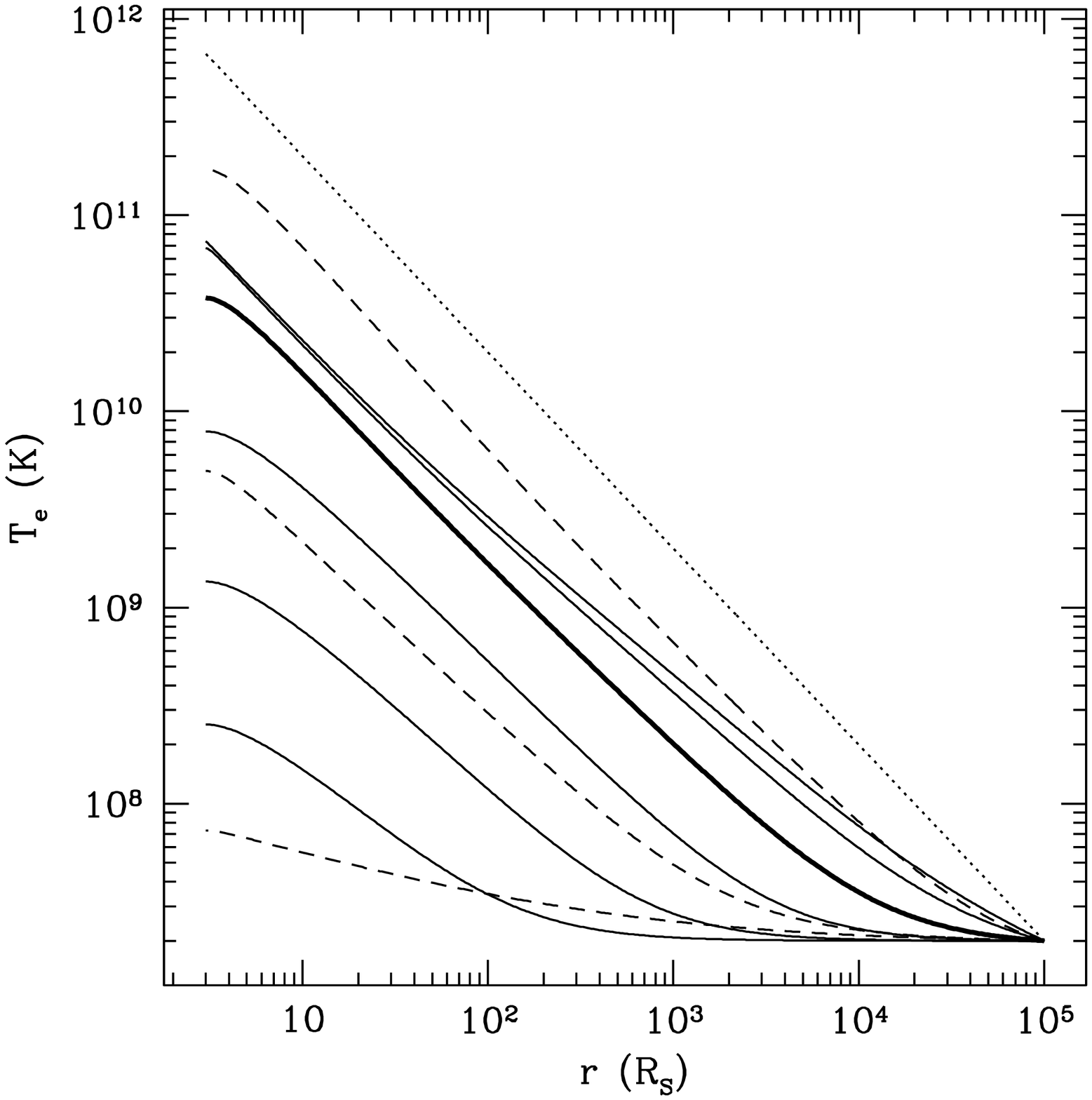}
\caption{Electron temperature with $n \propto r^{-0.5}$  for various
$\alpha_c/\alpha$ and $f_e$.  The thick solid line is the calculation
from Figure~\ref{f1} with $p = 0.5$, $\alpha_c/\alpha = 1$ and $f_e =
0.1$.  The solid curves correspond to $\alpha_c/\alpha$ varying from $0.01$ to 
$1000$ by factors of $10$ (from top to bottom).  
The dashed curves are for $f_e = 0.5$ (top),
$0.01$ and $0$ (bottom).  The dotted line is the
virial temperature $T_{vir} \propto r^{-1}$. Note that for this
density profile the solutions are never super-virial.}
\label{f2}
\end{figure}

\begin{figure}
\plotone{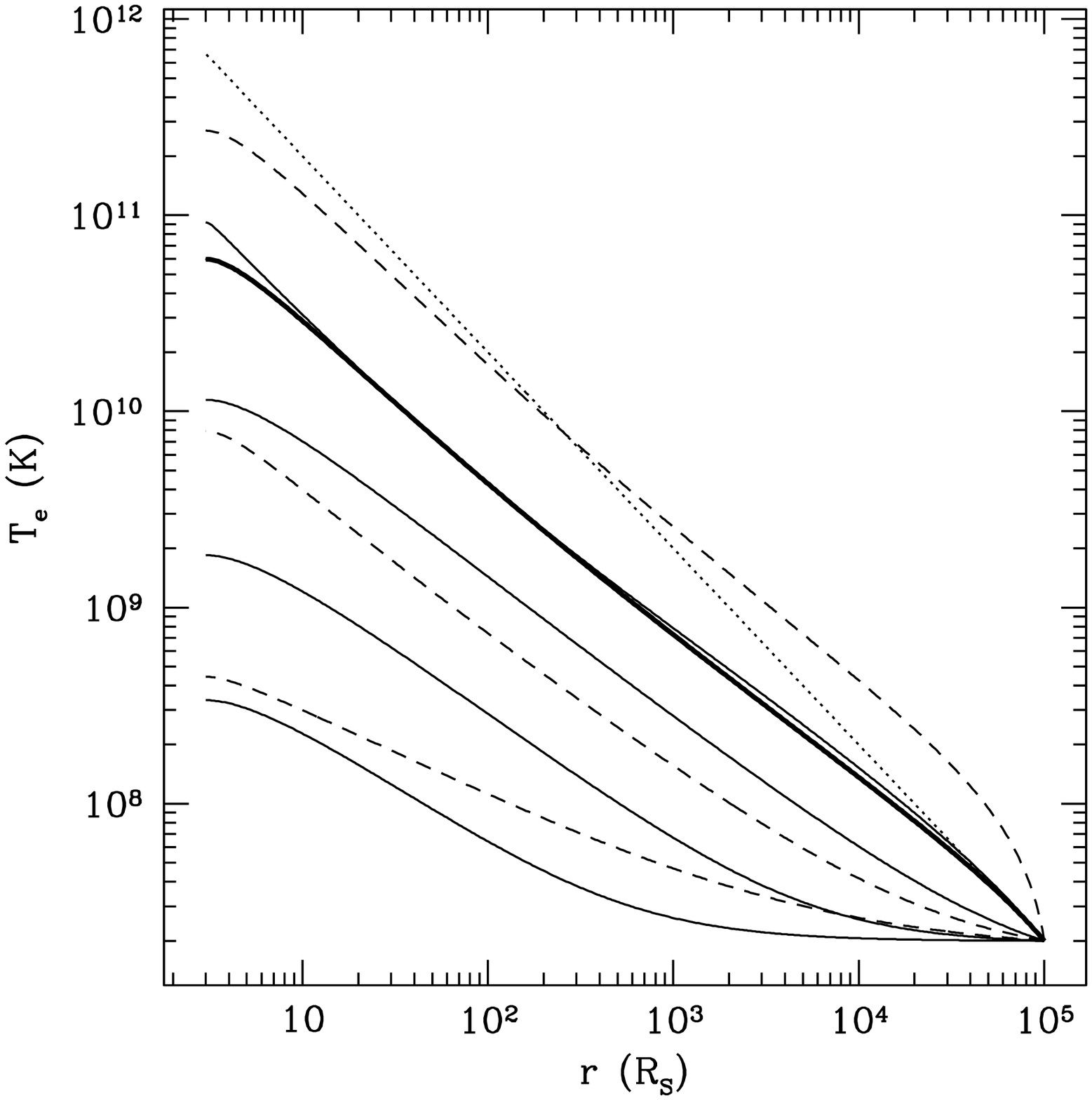}
\caption{Electron temperature with $n \propto r^{-1}$ for various
$\alpha_c/\alpha$ and $f_e$.  The thick solid line is the calculation
from Figure~\ref{f1} with $p = 1$, $\alpha_c/\alpha = 1$ and $f_e =
0.1$.  The solid curves correspond to $\alpha_c/\alpha$ varying from $0.1$ to 
$1000$ by factors of $10$ (from top to bottom).  
The dashed curves are for $f_e = 0.5$ (top),
$0.01$ and $0$ (bottom).  The dotted line is the
virial temperature $T_{vir} \propto r^{-1}$.}
\label{f3}
\end{figure}

We have shown that electron heat conduction can have a significant
effect on the properties of radiatively inefficient accretion flows
(RIAFs).  We considered two model problems to demonstrate this, both
of which assume spherical symmetry.  First, for the canonical Bondi
problem, we find that if a few percent of the accretion power is
transferred to large radii by heat conduction, the heat flux can
unbind the gas at large radii (eq. [\ref{trbondi}]).  Generalizing
this calculation to allow for a power-law density profile with $n(r)
\propto r^{-p}$ (\S \ref{FDP} and Figs. \ref{f1}-\ref{f3}), we solved
for the electron temperature profile in the presence of saturated heat
conduction; $p = 3/2$ is appropriate for the spherical Bondi problem,
while $p \sim 1/2-1$ is typically found in numerical simulations of
rotating radiatively inefficient accretion flows.  Our calculations
show that for density profiles with $p \gtrsim 1$, and for plausible
values of the electron conductivity and electron heating, the electron
temperature is super-virial at large radii (see
Fig.~\ref{f1}-\ref{f3}).  This indicates that the effects of
conduction on the dynamics of the accretion flow must be
self-consistently taken into account.

Motivated by this initial calculation, we solved the problem of
spherical accretion in the presence of heat conduction and electron
heating.  For appreciable electron heating rates ($f_e \sim 0.1$) and
dimensionless conductivities $\alpha_c \sim 0.1-1$ (see \S \ref{EE}
for the details of the electron energy equation), the accretion rate
is reduced by $\sim 1-3$ orders of magnitude relative to the canonical
Bondi rate (\S\ref{section:bondi}; Fig.~\ref{f8}).  Physically, this
is because some of the energy dissipated as heat is conducted to large
radii where the increase in pressure stifles the inflow of matter.
This decrease in $\dot M$ relative to the Bondi accretion rate is
consistent with observations of the Galactic Center, which indicate
that the accretion rate is $2-3$ orders of magnitude lower than the
Bondi rate \citep{qg00b,bower,marrone}.

As discussed in \S\ref{EE}, electron heat conduction is only important
at very low accretion rates $\lesssim 10^{-5}-10^{-6} \, \dot M_{\rm EDD}$,
when the cooling time of the electrons via synchrotron radiation is
longer than the conduction time in the plasma (eq. [\ref{MDUL}]).
Such accretion rates are appropriate for very low-luminosity AGN such
as Sgr A* in the Galactic Center, massive black holes in early type
galaxies (e.g., \citealt{TD}), and some X-ray binaries in quiescence.
For higher accretion rates, the electrons cool more rapidly than
energy can be redistributed by conduction, and so electron conduction
is unimportant. Ion heat conduction could in principle be relevant at
all accretion rates since the ion cooling time in RIAFs is always
longer than the inflow time (by construction); this is, however,
harder to quantify because the ion conduction time is at best
comparable to the inflow time of the plasma.

All of the calculations in this paper assume spherical symmetry.  This
is formally appropriate only in the absence of significant angular
momentum, when the circularization radius of the inflowing material is
small compared to the gravitational sphere of influence of the central
object.  We suspect, however, that the calculations presented here are
relevant to rotating accretion flows as well, given the importance of
energy transport in determining the structure of RIAFs (e.g.,
\citealt{bb99}).  In a rotating accretion disk, however, the conducted
energy may be transported to the surface of the disk, rather than to
large radii.  This conductive flux might contribute to driving a
thermal outflow. Indeed, the two-dimensional calculations of
\cite{tm06} indicate that conduction helps drive a bipolar outflow.
Numerical simulations of RIAFs with heat conduction are necessary to
further quantify the importance of conduction and to assess whether
any resulting heat flux is primarily radial or vertical.  Existing
simulations of RIAFs -- which do not include thermal conduction --
find density profiles with $p \approx 1/2-1$, in which case our
calculations indicate that conduction is unlikely to lead to
super-virial temperatures which would be dynamically important (see
Figs.~\ref{f1}-\ref{f3}).  Even in this case, however, conduction
would be important for determining the electron temperature profile
and thus the observed radiation.

Throughout this paper we have treated the dimensionless conductivity
$\alpha_c$ as a free parameter in order to explore the effects of
conduction in RIAFs.  A precise determination of $\alpha_c$ in this
context requires understanding the effective conductivity in a
magnetized, turbulent plasma.  Wave-particle scattering by high
frequency turbulence can limit the mean free path of particles in low
collisionality plasmas, and thus the conductivity.  In the Appendix,
we argue that instabilities generated by the nonlinear evolution of
the MRI in RIAFs lead to an upper limit on the ion conductivity of
$\alpha_c \lesssim 0.1$, in which case ion conductivity is unlikely to
be dynamically important.  We find that there is no corresponding
limit on the electron conductivity, although a large heat flux itself
may be self-limiting by generating whistler instabilities (e.g.,
Pistinner \& Eichler 1998).  Independent of wave-particle scattering,
electron transport perpendicular to a static magnetic field is
suppressed due to the small value of the electron Larmor radius.
However, the rapid separation of neighboring field lines in a
turbulent medium results in an effective isotropic conductivity
comparable to the field-free conductivity
\citep{jok73,nm01,cl04,cm04,mcb04}. This suggests that electron heat
conduction will by dynamically important even in the presence of a
tangled magnetic field, with $\alpha_c \sim 0.1-1$.  With this value
of the electron conductivity, we find that conduction significantly
modifies the dynamics of Bondi accretion (see Figs. \ref{f7}-\ref{f8}).

\begin{figure}
\plotone{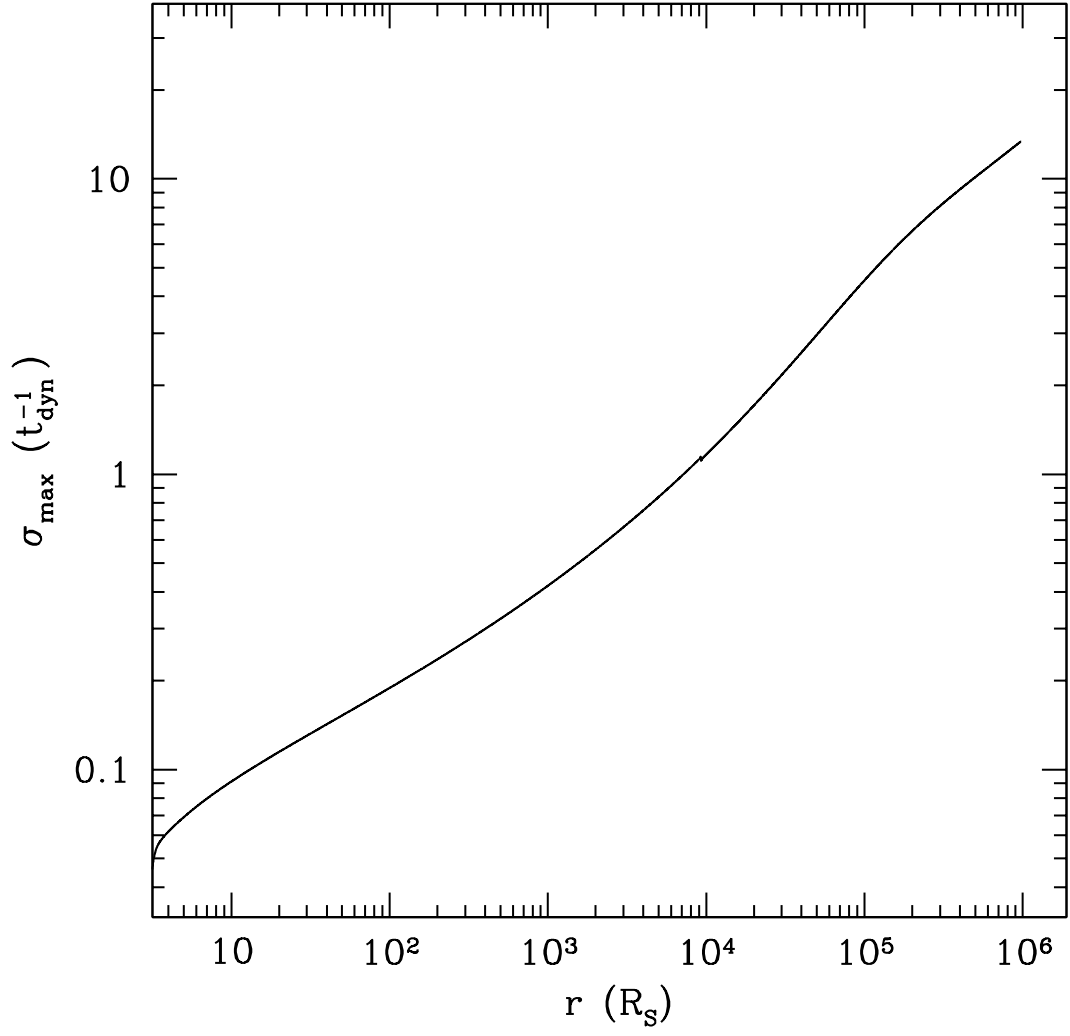}
\caption{MTI growth rate in units of the inverse infall time for a spherical accretion solution from \S
\ref{section:bondi}; $f_e = 0$ and $\alpha_c = 0.1$. Plotted as a function of radius is the maximum growth rate (equation [29] of \citealt{bal00}), achieved when the conduction time is short but the tension forces are small.}
\label{f9}
\end{figure}

\citet{bal00} has shown that low collisionality conducting plasmas are
unstable to convective-like motions in the presence of an outwardly
decreasing temperature gradient (the MTI).  In a non-rotating
atmosphere, \cite{ps05} demonstrate that the MTI leads to magnetic
field amplification and a substantial heat flux down the temperature
gradient.  It is natural to speculate that our solutions represent the
non-linear saturation of the MTI in accretion disks.  However,
simulations of the MTI in RIAFs are necessary to test this hypothesis
since it is not clear what effect differential rotation and the MRI
will have on the results of \cite{ps05}.  The MTI will also be
important in purely spherical accretion, as indicated by
Figure~\ref{f9}, which shows the growth rate of the MTI in units of
the inverse infall time calculated from one of our spherical accretion
solutions with $f_e = 0$ and $\alpha_c = 0.1$.  As Figure ~\ref{f9}
shows, the MTI will be particularly important at large radii and will
likely significantly modify existing results on the impact of magnetic
fields on spherical accretion (e.g., those of \citealt{in02}).

Although our focus in this paper is on the electron energetics in
RIAFs, we have also carried out some preliminary calculations
including the ions (neglecting ion conduction for reasons discussed
above and in \S \ref{EE}).  For spherical accretion with significant
ion heating, the accretion rate is reduced relative to the Bondi rate
by even more than is shown in our electron-only calculations in Figure
\ref{f8}.  This is consistent with the arguments in \S
\ref{section:bondi}. The ions are non-relativistic throughout most of
the flow, and therefore have $\gamma_i \simeq 5/3$. In the presence of
ion heating, the effective adiabatic index of the flow is $\gtrsim
5/3$, in which case there is no transonic Bondi solution in Newtonian
gravity.  Instead, the inflowing matter is in approximate hydrostatic
equilibrium throughout the bulk of the flow and the solution never
achieves the free-fall density profile of $\rho \propto r^{-3/2}$ (see
the discussion around eq.  [\ref{GAMC}]). The further reduction in 
accretion rate due to the inclusion of ion heating (even absent ion 
conduction), while strengthening our conclusions, is a consequence 
of the singular nature of the $\gamma_i \simeq  5/3$ Bondi solution, for 
which $\rS \simeq 0$. Including additional physical effects such as rotation 
will partially remove this singular behavior and may be required to 
realistically assess the impact of ion heating and conduction on the 
dynamics of hot accretion flows.

As noted in the Introduction, radiation emitted from close to the
accreting object can play a role analogous to that of conduction in
heating up the outer region of the accretion flow to super-virial
temperatures. This mechanism has been studied in the context of RIAFs
by \cite{po98,po99,po01}. The accretion flow in that case can exhibit
time dependence due to the lack of a steady-state solution or to thermal
instability \citep{owym76,cos78,bb80,kl83}. We do not expect time
dependence in our problem, since our calculations show that a
steady-state solution does exist and conduction tends to stabilize
thermal instabilities \citep{kl83}; a time-dependent calculation or a
stability analysis would, however, be required to confirm this.  We
also note that our calculations are only relevant for systems with
sufficiently low luminosities as to be unaffected by radiative
preheating (as discussed above, our solutions require low luminosities
so that a significant fraction of the accretion energy can be
conducted outwards rather than being radiated away).

\acknowledgements

We thank Steve Balbus, Greg Hammett, Kristen Menou, Prateek Sharma,
and the referee, Jerry Ostriker, for useful conversations and
comments.  This work was supported in part by NSF grant AST 0206006,
NASA grant NAG5-12043, an Alfred P. Sloan Fellowship, and the David
and Lucile Packard Foundation.

\newpage

\begin{appendix}

\section{An Upper Limit on $\alpha_c$}

In a collisionless plasma, the temperature and pressure need not be
the same perpendicular ($T_\perp$) and parallel ($T_\Par$) to the
local magnetic field.  The temperature anisotropy cannot, however, be
arbitrarily large because high frequency waves and kinetic
microinstabilities feed on the free energy in the temperature
anisotropy, effectively providing an enhanced rate of collisions that
isotropizes the pressure tensor via pitch angle scattering.  This in
turn limits the mean free-path of particles and thus the heat
conductivity.  These issues are discussed in detail in \cite{sharma}
in the context of nonlinear simulations of the MRI in a collisionless
plasma. Here we use this effect to estimate an upper limit on
$\alpha_c$ for ions and electrons in RIAFs.  For concreteness we focus
on instabilities generated by $T_\perp > T_\Par$.  For conditions
appropriate to RIAFs, the most important instability regulating the
ion (electron) pressure anisotropy is the ion cyclotron (whistler)
instability.\footnote{The mirror instability provides a comparable
rate of ion pitch angle scattering \citep{sharma}.}

For electromagnetic fluctuations slow compared to the cyclotron
frequency, the magnetic moment ($\mu \propto T_\perp/B$) is an
adiabatic invariant.  As a result, temperature anisotropies with
$T_\perp \ne T_\Par$ are created due to the fluctuating magnetic field
associated with the MRI.  Formally, a plasma with any nonzero
temperature anisotropy can be unstable to the ion cyclotron and
whistler instabilities \citep{sti92}.  However, there is an effective
anisotropy threshold given by the requirement that the unstable modes
grow on a timescale comparable to the disk rotation period. Gary and
collaborators have analyzed the ion cyclotron instability in detail
through linear analysis and numerical simulations. \citet{gar97} and
\citet{gar94} calculate the anisotropy required for a given growth
rate $\gamma$ relative to the ion cyclotron frequency $\Omega_i$ \be
\label{eq:gary_thresh}
\frac{p_\perp}{p_\Par} - 1 > \frac{S}{\beta_{\Par,i}^q} \ee where
$\beta_{\Par,i} = p_{\Par,i}/(B^2/8\pi)$ and where $S=0.35$ and
$q=0.42$ are fitting parameters quoted in equation (2) of
\citet{gar94} for $\gamma/\Omega_i=10^{-4}$. Moreover, for $\gamma \ll
\Omega_i$ the threshold anisotropy depends only very weakly on the
growth rate $\gamma$.  As a result, equation (\ref{eq:gary_thresh})
provides a reasonable estimate of the pressure anisotropy required for
pitch angle scattering by the ion cyclotron instability to be
important on a timescale comparable to the disk rotation period
($\Omega^{-1}$).

If the characteristic timescale for the magnetic field to change by
order unity due to turbulence in the accretion disk is $\Omega^{-1}$,
then the rate of pitch angle scattering (the ``collision frequency'')
required to ensure that the pressure anisotropy never exceeds the
threshold given by equation (\ref{eq:gary_thresh}) is $\nu \delta p
\approx \Omega p \rightarrow \nu \approx \Omega \beta_{\Par,i}^q/S$
where $\delta p = p_\perp - p_\Par$ is the pressure anisotropy at the
threshold in equation (\ref{eq:gary_thresh}).  For pitch angle
scattering at a rate $\nu$, the ion conductivity is given by \be
\kappa_i \approx {n v_i^2 \over \nu} \approx n v_i r \left(H \over
r\right) \left(S \over \beta_{\Par,i}^q\right)
\label{ionlimit} \ee where the scale-height of the disk is given by 
$H \approx v_i/\Omega$.  Equation (\ref{ionlimit}) corresponds to
$\alpha_c \approx S \beta_{\Par,i}^{-q} H/r$.  Nonlinear simulations
of the MRI in RIAFs imply $\beta \sim 10$ in the bulk of the disk, in
which case $\alpha_c \approx 0.1 H/r$ for the ions.  This is, of
course, only an upper limit because there may be additional sources of
wave-particle scattering and because equation (\ref{ionlimit}) is the
conductivity along field lines while the conductivity used throughout
the rest of this paper is an effective isotropic conductivity.

The argument for an upper limit on the electron conductivity proceeds
analogous to that above, with equation (\ref{eq:gary_thresh}) being
replaced by the corresponding limit on the electron anisotropy induced
by the whistler instability.  For non-relativistic electrons,
\cite{gar96} show that the threshold for the whistler instability is
similar to that of equation (\ref{eq:gary_thresh}) with
$\beta_{\Par,i} \rightarrow \beta_{\Par,e}$.  Thus the electron
conductivity is given by \be \kappa_e \approx {n v_c^2 \over \nu}
\approx n v_c r \left(H \over r\right) \left(S \over
\beta_{\Par,i}^q\right) \left(T_i \over T_e\right)^q \left(v_c \over
v_i \right). \label{elimit} \ee Equation (\ref{elimit}) is far less
restrictive than equation (\ref{ionlimit}), generally implying
$\alpha_c \gtrsim 1$ for the electrons, i.e., no limit on $\alpha_c$
at all (saturation of the electron heat flux, however, implies $\alpha_c
\lesssim 1$).  This conclusion strictly only applies when the
electrons are non-relativistic and more work is needed to understand
the extension of this argument to relativistic electrons.

\end{appendix}

\newpage

\end{document}